# Preprint Clinical Feedback and Technology Selection of Game Based Dysphonic Rehabilitation Tool

Zhihan Lv
FIVAN, Valencia, Spain
Email: lvzhihan@gmail.com

Chantal Esteve
FIVAN, Valencia, Spain
Email: chantal@fivan.org

Javier Chirivella
FIVAN, Valencia, Spain

Pablo Gagliardo
FIVAN, Valencia, Spain
Email: pablog@fivan.org

*Abstract*—An assistive training tool software for rehabilitation of dysphonic patients is evaluated according to the practical clinical feedback from the treatments. One stroke sufferer and one parkinson sufferer have provided earnest suggestions for the improvement of our tool software. The assistive tool employs a serious game as the attractive logic part, and running on the tablet with normal microphone as input device. Seven pitch estimation algorithms have been evaluated and compared with selected patients voice database. A series of benchmarks have been generated during the evaluation process for technology selection.

## I. Introduction

Using video game as the vision feedback has been already demonstrated as a useful HCI approach for rehabilitation assistive method [10]. Some previous research have proved the usability of the game in speech therapy [28], for children with autism [5], and parkinsons disease patients [6]. However, none of these researches has considered the success rate of recognition of patients' voice by different pitch estimation algorithms from the view of computer technology.

In this research, we evaluated seven pitch estimation algorithms using our patient voice database, and then gave a preliminary benchmark of the performance of pitch recognition of patient voice. Video game is employed as vision feedback tool to evaluate the efficacy of the patient voice exercise, which can be considered as the long-time continued evaluation of the severity of dysphonia. Pitch is detected as the evaluation factor which is also the controlable input of the games. We define our research as a biometric information management system design, in which the biometric refer to voice pitch specifically, while serious games are employed as the user interface. In our previous research [13], we mentioned that our design science based software development just go forward a step, and thus 'the next step is to iteratively improve it and to develop experiments to test it before eventually moving on to conducting clinical tests as the assistive tool for rehabilitation of dysphonic patients'. Therefore, we recorded and assessed the feedback from the treatments for improving the assistive tool. In other side, we supposed the therapist could supervise and train the patient side by side since the microphone arrays of kinect is enabled to filter ambient noise. In this paper, we discussed this idea according to the practical clinical application as well as the direct feedback from the treatment.

## II. Clinical feedback from treatment

In clinical, dysphonia is measured using a variety of tools that allow the clinician to see the pattern of vibration of the vocal folds. The amount of dysphonic patients is much smaller than others limbs disable patients. In our sample database, it's one fourth (fifteen of sixty), in which four or five are expected to be recovered with computer games as assistive tools. According to our sample database, the dysphonic voice can be hoarse or excessively breathy, harsh or rough, but some kind of phonation is still possible, therefore, the clear speech recognition is not expected. The patients usually have inabilities of pronunciation of the high pitches. The hopeful assistive tool is able to retrieve the quantified long-time voice rehabilitation information.

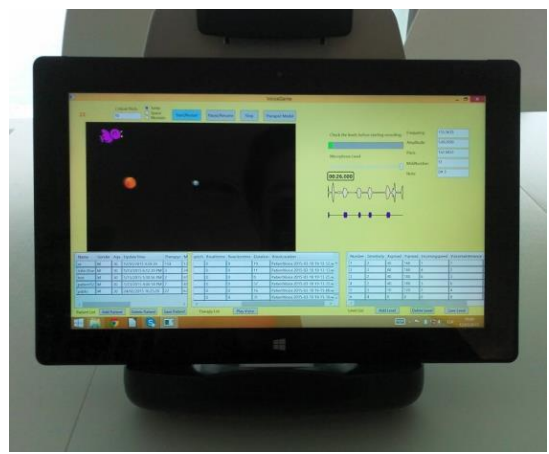

Fig. 1. New user interface of the proposed tool which is suitable for tablet with normal micphone.

Previously, we have discussed with a rich-experience therapist regarding general exercise process in order to identify and define the problem pragmatically. The first version of the assistive tool is designed according to the therapist's suggestions and considering 'Intuitive', 'Entertainment', 'Incentive' as main design factors. The latest version of the software proposed in this paper supports common mic and gets rid of Kinect2. Because we have realized that the treatments are like the driving training, the independent driving isn't expected at the early training usually. Especially when the game is out of control but the patient cannot increase the pitch at all, the therapist should rapidly make voice and help the patient to complete the game, which gives patient a paradigm and can reduce anxiety and boost confidence of the patients. The new user interface is shown as in figure 1. The left side of the UI is the game logic, the right side is the estimated pitch values monitor which presents frequency, pitch, amplitude, midi note, midi note number, sample, duration.

The tool is evaluated in the clinical treatment for three patients. The first patient is a stroke sufferer who always sits inside the wheelchair. The patient cannot speak, but he can still make voice. After exercise four times, the patient control the avatar in the game successfully. The therapist suggested that the distance between the obstacles in the game should be shorter so that the patient could control the avatar to easily evade the obstacles as they want. During the game, we can see the patient struggle to talk and ask the questions about the games, but he cannot make voice clearly. The stroke patient was evaluated again at the next day. After we modified the density of the planets and the critical value of pitch, the patient can play with the game well, since the game it's much easier.

The second patient is a parkinson sufferer who can sit and stand independently. The therapists played the game to instruct the game rule at the beginning and then the patient tried to play with the game. After the patient repeated the game four times, the patient complete the game successfully. The patient's voice is not clear, so the pitch estimation algorithm cannot detect it. A more sensitive algorithm which is suitable for patient's voice is needed. The critical values of the player pitch for patients are usually vastly different from the normal people. We set the critical value as half as that for the normal people at the beginning, but it's still impossible for patients to reach. We modified the setting iteratively. It doesn't work until the value of the setting is eight times easier than normal people. The patient is unable to change the pitch, so we try to use extreme low pitch ($50Mel$) as the critical pitch.

The therapist wish a bigger game logic user interface (UI), because the patient's vision is not enough good to see the whole scene, another reason is that the movable space for the avatar is too limited. The current avatar in the game is also too small to manipulate. Similarly, the patient thinks the obstacles are also too small, he cannot see them very clearly. We have observed that the patient struggled to try to play with the game UI by touch during the game, which is because of that the current interaction approach is not idiomatic.

The most intolerable problem is that the algorithm is still not sensitive enough for patients voice. Sometimes, the detection of some pitches are missing. During the game, we noticed that different voice or different timbre got different success rate of detection. For example, 'Wo' was recognized with higher success rate than 'Aa'. The detection of 'Yi' had a very good performance. By making voice of 'Yi', the patient passed the easiest level of the game. The detection of 'Ai' did't work in our game.

After twice treatment for each patient, the therapist start to supervise the patients to play with the game without our attending. After three times treatments in one week, the patients can control the game by their pitch very well, even they can make the voice of 'Ai' effortlessly. Because they have understood the game rule now and distinguished the pitch and loudness. Previously, the patients was struggling to make voice but actually they only increased the loudness which was not the controller factor of the designed game. Now the patients don't put forth their strength to make voice any more, instead, they let their voice be slightly and increase the pitch skillfully, as the therapist teach them, as well as the experience that they realized from the game. The easiest levels of the games are already suitable for both patients. We are going to improve both UI and pitch estimation algorithm of the software as the therapist suggested so that it can be suitable for most of the dysphonic patients.

The situation of the third patient is better than the previous two patients. The therapist tried to treat this patient using senior level of the designed game. However, the result indicated the senior level was still too difficult for the patient even if he has better situation. The main reason is the lack of the sensitivity of the pitch estimation algorithm. The minor reason is that the great disparity between normal people and dysphonic patients.

### III. ALGORITHM COMPARISON

As we have mentioned, in our previous work [13] the frequency of the patient voice is identified by FFT (Fast Fourier Transform) based frequency-domain approach which turns the voice wave into a frequency distribution. Further more, the pitch (tone) and midi number are calculated depend on frequency. The therapist evaluated the pitch estimation approach with piano and proved the reliability and accuracy of it. We have considered the voice characteristics like non-continuity and low-loudness of dysphonic patients, so we concluded that the sensitivity and suitability of the pitch extracting method are the decisive factors beside the accuracy rate. In the clinical feedback section, we have introduced that the current FFT based pitch estimation algorithm can only detect some certain pitches. Therefore, there is a practical need that creating or finding a stable and sensitive enough pitch estimation algorithm for the special voice of patients in different conditions.

In this paper, we will evaluate seven pitch estimation algorithms, which are also known as pitch detection or pitch recognize algorithms. The seven pitch estimation algorithms are as followed: AdvancedAutoCorrelator [1], Dynamic Wavelet Algorithm [7] which is implemented at [25], FastYin [22] which is a faster implementation of Yin [4], MPM [23] by Philip McLeod and Geoff Wyvill, Yin [4] which is implemented at [2], FFT [24] which is implemented at [33], classical AutoCorrelator. The implementation is based on C# with WPF, the algorithms refer to the source code of TarsosDSP [26].

We filtered the pitch above $400Mel$ and only keep the pitch below $400Mel$ which is the limit of patients voice, because sometimes super high pitches ($> 2000Mel$) were detected.

In the first step, we compared the performance and time-consuming of seven algorithms with different buffer size (1024, 2048, 4096, 8192, 16348) using sine wave as input. The sine wave is a mathematical curve that describes a smooth repetitive oscillation, also on behalf of the analog signal. Figure 2 left indicates that with the amplification of the buffer size, the curve tends to smooth.

For example, when buffer size is 1024, Dynamic Wavelet, Yin and McLeod have error from midi note number of 35 to 40. FastYin has error from 35 to 45. FFT and AdvanceAutocorrelator have regular errors during the whole range. Autocorrelator has irregular error during the whole rage.

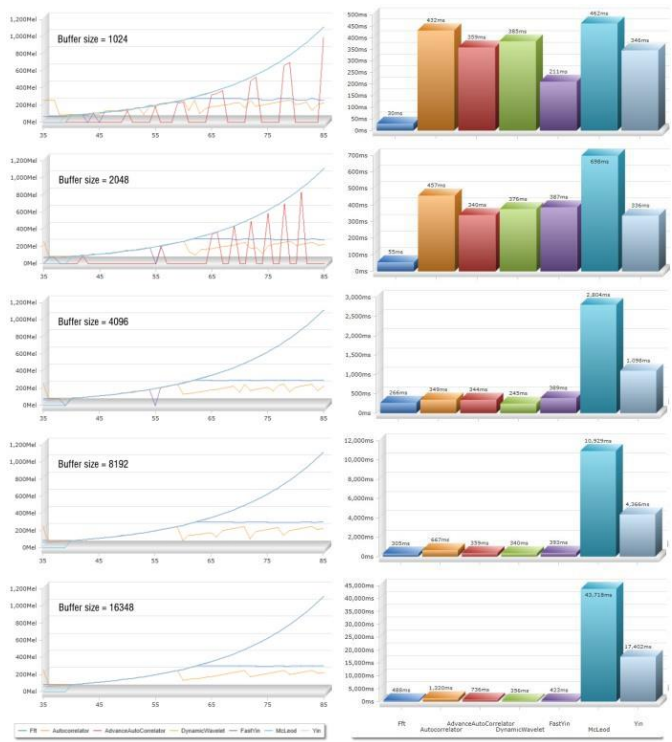

Fig. 2. The performance comparison of the seven algorithms with different buffer size (1024, 2048, 4096, 8192, 16348). Left: Pitch estimation from sine wave, the x-axis represents midi note number; Right: Time consuming of algorithms.

When buffer size is 4096, AdvanceAutocorrelator, Dynamic Wavelet, Yin have not any error in the whole range. FastYin and Mcleod have errors at 40, 55 and 35 to 40 respectively. Fft and Autocorrelator have errors at 35 to 60 and 36 to 60 respectively.

When buffer size is 16384, only Fft and Autocorrelator have serious errors after 60. Mcleod has tiny error before 40. Besides, all the other algorithms have normal curve.

Figure 2 right indicates that the time consuming of Mcleod increases immensely with the increase of buffer size, while the time consuming of other algorithms aren't changed significantly.

In the second step, the algorithms are compared using patient voice database. Figure 3 shows the performance comparison of the algorithms with 4096 buffer size using parkinson patient voice record. This voice section recorded a few of voice of 'Wo'. Mcleod cannot correctly estimate the pitch of patient voice when the buffer size is more than 1024 so we won't add it in this figure. As shown in Figure 3, all the estimation results of Autocorrrlator and Fft are non-zero values, which means successful estimated pitch. Yin and FastYin have only one estimated non-zero pitch value, which is too limited for pitch controlable game. AdvanceAutocorrelator and Dynamic Wavelet get median intensive pitches between the two kinds. The comparison results indicates that Autocorrelator algorithm gets the most intensive pitches which means it is the most sensitive pitch estimation algorithm for patient voice. No zero value pitches estimated also reveals that Autocorrelator and Fft

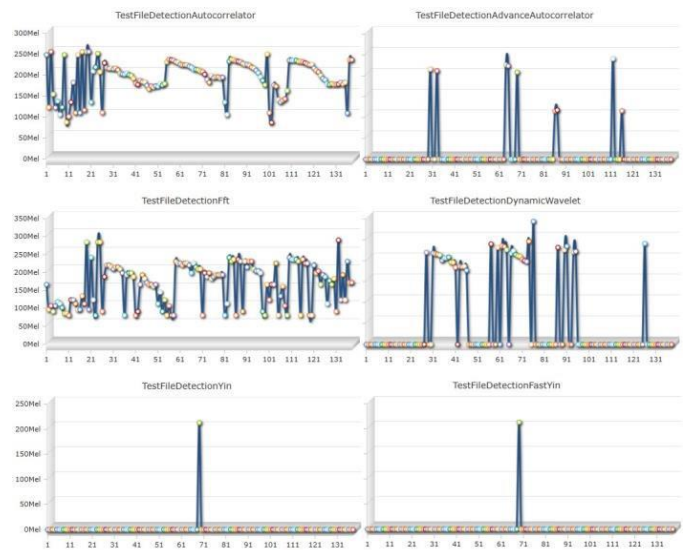

Fig. 3. The performance comparison of the seven algorithms with 4096 buffer size using parkinson patient voice record, the x-axis represents midi note number. This voice section recorded a few of voice of 'Wo'.

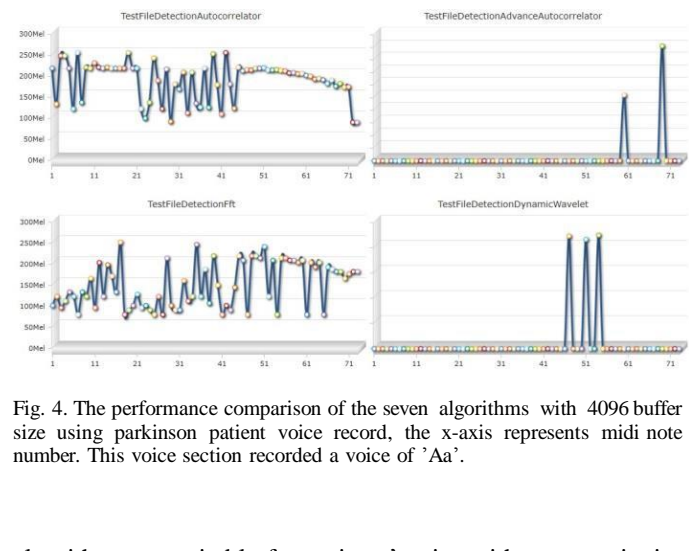

Fig. 4. The performance comparison of the seven algorithms with 4096 buffer size using parkinson patient voice record, the x-axis represents midi note number. This voice section recorded a voice of 'Aa'.

algorithms are suitable for patients' voice with non-continuity characteristic.

The flaws of Autocorrelator and Fft are that the pitch estimation errors happen when the midi note numbers are more than 60, which means the two algorithms are not suitable for high-pitched voice. Fortunately, the limit of patient voice pitch is 60 as we mentioned in our previous work [13], so the limit of the two algorithms are in the scope of the patients' voice. Therefore, Autocorrelator is the most suitable candidate algorithm for patient voice pitch estimation.

Figure 4 presents a pitch estimation of a piece of voice section which records a voice of 'Aa' by the same parkinson patient. This voice section recorded a voice of 'Aa'. The pitch estimated by seven algorithms with 4096 buffer size. Yin and FastYin cannot estimate any pitch from this piece of voice. The observation result of the graph also proves that Autocorrelator is the most sensitive algorithm and suitable for patients' voice with non-continuity characteristic.

## IV. Conclusion

The contributions of this paper include two points. The first one is the evaluation of the designed assistive tool for dysphonic rehabilitation. One stroke sufferer and one parkinson sufferer have brought us their practical suggestions according to their experiences to improve the software in the treatments. The second point is the most suitable pitch estimation algorithm selection. In order to select the best algorithm, seven algorithms have been evaluated and compared with selected patient voice database. Even through the comparison results indicate that Autocorrelator is the most sensitive algorithm and suitable for patients voice with non-continuity characteristic, a custom pitch estimation algorithm which can refine patient voice is still expected.

Virtual reality has been widely applied in many fields range from everyday entertainment and communication [12] [34] to traditional research fields such as geography [21] [29], ocean [27] [19], city [18] [9] [8] and biology [30] [20], clinical assist [13] [14] [17]. Some novel interaction approaches are considered to be integrated in our future work [11] [16] [15]. The new network data management algorithm [31] [32] , smart grid system [3] [35],

## Acknowledgment

The authors would like to thank Sonia Blasco, Vicente Penades and Oktawia Karkoszka. The work is supported by LanPercept, a Marie Curie ITN funded through the 7th EU Framework Programme(316748).